\documentclass[12pt]{article}
\usepackage{fullpage}
\usepackage{graphicx}
\newcommand{\be}{\begin{equation}}
\newcommand{\ee}{\end{equation}}

\def\w {\omega}

\newcommand{\news}{\setcounter{equation}{0}\quad}
\def\ben{\begin{equation}}
\def\een{\end{equation}}
\def\bea{\begin{eqnarray}}
\def\eea{\end{eqnarray}}
\begin{document}
\title{
\begin{flushright}\ \vskip -2cm {\normalsize{\em DCPT-08/57 }}\end{flushright}
\vskip 2cm Q-balls, Integrability and Duality}
\author{Peter Bowcock, David Foster and Paul Sutcliffe\\[10pt]
{\em \normalsize Department of Mathematical Sciences,
Durham University, Durham DH1 3LE, U.K.}\\[10pt]
{\normalsize Emails: peter.bowcock@durham.ac.uk, \quad d.j.foster@durham.ac.uk, \quad p.m.sutcliffe@durham.ac.uk}
}
\date{December 2008}
\maketitle
\begin{abstract}
This paper is concerned with the dynamics and interactions of Q-balls
in (1+1)-dimensions. The asymptotic force between well-separated Q-balls is
calculated to show that Q-balls can be attractive or repulsive depending
upon their relative internal phase. An integrable model with exact 
multi-Q-ball solutions is investigated and found to be of use in explaining
the dynamics in non-integrable theories. In particular, it is demonstrated
that the dynamics of small Q-balls in a generic class of non-integrable models
tends towards integrable dynamics as the charge decreases.
Long-lived oscillations of a single Q-ball can also be understood in terms
of a deformation of an exact breather solution in the integrable model.
Finally, we show that any theory with Q-ball solutions has a dual description
in which a stationary Q-ball is dual to a static kink, with an interchange
of Noether and topological charges.     
\end{abstract}

\newpage
\section{Introduction}\news
Q-balls are time-dependent non-topological solitons which 
carry Noether charge associated with a global $U(1)$ symmetry of
a nonlinear field theory \cite{Co}.  
Although they arise in a variety of theories
perhaps the most important example is the minimal supersymmetric standard
model, where the Noether charge is associated with the $U(1)$ symmetries
of baryon and lepton number conservation and the relevant fields 
correspond to squark or slepton particles \cite{K2}. In this scenario
Q-balls are condensates of a large number of squarks or sleptons and could
play a role in baryogenesis through the Affleck-Dine mechanism \cite{AD}.
Potential interesting cosmological consequences include contributions
to dark matter and isocurvature baryon fluctuations.

In order to fully investigate the properties of Q-balls a crucial
ingredient is to understand Q-ball dynamics and multi-Q-ball interactions.
This is not a simple task as Q-balls have time-dependent internal phases
and are solutions of nonlinear field theories. 

There is a vast literature on stationary Q-balls (see the recent 
work \cite{TCS} and references therein) but only limited studies on
dynamical aspects of Q-balls, mainly involving numerical simulations
 \cite{BS,AKPF}. These numerical investigations
reveal complicated phenomena including phase-dependent forces,
charge exchange and Q-ball fission and fusion. These results are qualitatively 
similar for a range of theories and spacetime dimensions, 
including (1+1)-dimensions. Despite this fact, there is very little analytic 
understanding of Q-ball dynamics in any system.  

In this paper we address this issue by performing some of the first
analytic studies of Q-ball dynamics.
We study Q-balls in (1+1)-dimensions and compare our findings with numerical
results. We calculate the force between well-separated Q-balls, verifying 
that the force can be either attractive or repulsive depending upon 
the relative internal phase between the Q-balls. 
We make extensive use of an integrable
model for Q-ball dynamics to understand Q-ball interactions and perturbations
in non-integrable theories. Furthermore, it is demonstrated that the 
dynamics of small Q-balls in a general class of non-integrable theories is
 captured
by the integrable theory, with increasing accuracy as the charge decreases.
Finally, we describe how any theory with Q-balls has a dual description
in which a stationary Q-ball is dual to a static kink, with an interchange
of Noether and topological charges. 

Part of our study involves a generalization of Q-balls to theories
with a non-standard kinetic term. It is pointed out that a certain 
class of such theories is more conducive to Q-ball existence than standard
theories, as most of the usual constraints on the potential are removed. 
This is true in all spacetime dimensions and therefore suggests new 
possibilities for Q-balls in systems previously not considered.
Furthermore, non-standard kinetic terms are currently popular in
cosmological theories and string theory, so Q-balls may have applications
in these contexts.

\section{Q-balls}\news
In this section we first review some material regarding Q-balls in
theories that are typically used as the simplest examples possessing
Q-ball solutions. We then turn to a discussion of modified theories
with a non-standard kinetic term and explain why Q-balls are more generic
in a class of theories of this type, since most of 
the usual restrictions imposed on the potential are no longer necessary.

\subsection{Q-balls in standard theories}
Q-balls arise in various types of field theories, but the simplest example
consists of single complex scalar field with a global $U(1)$ symmetry.
Explicitly, consider the (1+1)-dimensional theory with Lagrangian density
\be
{\cal L}=\partial_\mu\phi\partial^\mu\bar\phi-U(|\phi|)\,,
\label{usuallag}
\ee
where, without loss of generality, we have chosen to set $U(0)=0.$ 

The Noether charge associated with the global $U(1)$ symmetry 
is given by
\be
Q=i\int_{-\infty}^{\infty} (\phi\partial_t\bar\phi-\bar\phi\partial_t\phi)\,dx.
\label{q}
\ee
A Q-ball positioned at the origin is a stationary solution of the form 
\be
\phi=e^{i\omega t}f(x),
\label{qball}
\ee
where $\w>0$ is the internal rotation frequency
and $f(x)$ is a real profile function satisfying the boundary conditions
$\frac{df}{dx}(0)=0$ and $f\rightarrow 0$ as $|x|\rightarrow\infty.$

Of course, given the stationary solution (\ref{qball}) the position can
be shifted by a spatial translation and the Q-ball can be
given an arbitrary velocity up to the speed of light 
(which is one in our units) by applying a Lorentz transformation.

Using the stationary ansatz (\ref{qball}) the Noether charge
takes the form
\be
Q=2\w \int_{-\infty}^{\infty} f^2\,dx.
\ee

The classical field equation which follows from the Lagrangian density
(\ref{usuallag}) is
\be
\partial_\mu\partial^\mu\phi+\frac{\partial U}{\partial \bar\phi}=0,
\label{uft}
\ee
which has a solution of the form (\ref{qball}) providing the
profile function obeys the ordinary differential equation
\be
\frac{d^2f}{dx^2}=-\omega^2 f+\frac{1}{2}\frac{dU}{df}.
\label{genfpp}\ee
Linearizing (\ref{genfpp}) and requiring a decaying
solution at spatial infinity implies an upper
limit on the frequency $\w$ given by
\be
\w^2<\w_+^2=\frac{1}{2}\frac{d^2U}{df^2}\bigg|_{f=0}
\label{con1}
\ee
Integrating (\ref{genfpp}) once, and using the boundary condition
at infinity, gives the first order equation
\be
\left(\frac{df}{dx}\right)^2=-\w^2f^2+U.
\label{bog}
\ee
The boundary condition that the derivative vanishes at $x=0$ requires 
that $f(0)\equiv f_0>0$
is a solution of the equation
\be
U(f_0)=\w^2f_0^2.
\label{con2}
\ee
This equation places a constraint on the type of potential $U(f)$
that allows Q-balls, and furthermore, for an allowed potential it determines a
lower limit on the frequency $\w.$ This is traditionally expressed
in the form
\be
\w^2>\w_-^2=\mbox{min}\left(\frac{U(f)}{f^2}\right),
\ee   
but this assumes that for all $f\ge 0$ then $U(f)\ge 0,$ that is $f=0$ is 
the global minimum of the potential. If $U(f)<0$ for any $f>0$
then clearly once the condition (\ref{con1}) is satisfied then 
the additional condition (\ref{con2}) places no further constraint on the
frequency $\w$ beyond the original assumption that $\w>0.$ Obviously, if 
$U(f)< 0$ for some $f>0$ then it is to be expected that some Q-ball
solutions will be unstable, since they are nonlinear excitations built
upon the vacuum $\phi=0,$ which will be a false vacuum.

For a potential which is polynomial in $|\phi|^2,$ the simplest form
allowing Q-ball solutions is the general $\phi^6$ potential 
\be
U(f)=f^2-f^4+\beta f^6,
\label{usualu}
\ee
where, without loss of generality, the coefficients in front of the first two 
terms have been scaled to unity using the freedom to rescale the spacetime
coordinates and the field. In the above potential $\beta$ is a non-negative 
parameter of the theory.
Writing (\ref{usualu}) in the form
\be
U(f)=f^2(1-\frac{1}{2}f^2)^2+(\beta-\frac{1}{4})f^6.
\ee
makes it clear that if $\beta>\frac{1}{4}$ then $f=0$ is the unique global 
minimum of the potential.
If $\beta=\frac{1}{4}$ then there are degenerate vacua
at $f=0$ and $f=\sqrt{2}.$ Q-balls in theories with degenerate vacua
were first considered in \cite{Sp}.
Finally, if $\beta<\frac{1}{4}$ then the potential is unbounded 
from below, so $f=0$ is a false vacuum. However, even in
this case $f=0$ is still a local minimum, so there will be
Q-ball solutions built upon the false vacuum.

With the general $\phi^6$ potential (\ref{usualu})
the nonlinear field equation (\ref{uft}) becomes
\be
\partial_\mu\partial^\mu\phi+\phi(1-2|\phi|+3\beta|\phi|^2)=0,
\ee
and the profile function equation (\ref{genfpp}) is
\be
\frac{d^2f}{dx^2}=-\omega^2 f+f(1-2f+3\beta f^2),
\ee
with solution
\be
f=\frac{\omega'\sqrt{2}}{\sqrt{1+\sqrt{1-4\beta \omega'^2}\cosh(2\omega'x)}},
\ee
where we have defined the complementary frequency $\omega'=\sqrt{1-\omega^2}.$

Using this exact solution the charge and energy can be calculated explicitly
to be 
\be
Q=\frac{4\omega}{\sqrt{\beta}}\tanh^{-1}\left(\frac{1-\sqrt{1-4\beta\omega'^2}}
{2\omega'\sqrt{\beta}}\right),
\ee
\be
E=\int_{-\infty}^{\infty} \left(\frac{df}{dx}\right)^2+\omega^2 f^2+U(f)
\,\,dx
=
\frac{4\w\w'+Q(4\beta(1+\w^2)-1)}{8\w\beta}.
\ee

\begin{figure}[ht]
\begin{center}
\includegraphics[width=12cm]{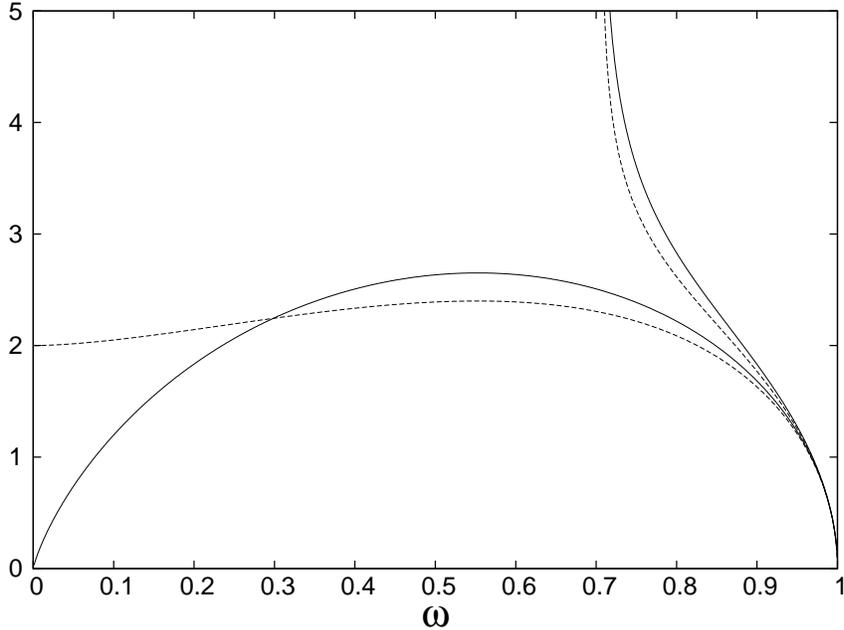}
\caption{
The charge $Q$ (solid curves) and energy $E$ (dashed curves)
as a function of $\w$ for $\beta=1/2$ (upper curves)
and $\beta=1/4$ (lower curves).
}
\label{fig-QE}
\end{center}
\end{figure}

These explicit expressions reveal that the existence of large Q-balls,
that is Q-balls with arbitrarily large values of the Noether charge $Q,$ 
depends crucially upon the parameter $\beta.$ Large Q-balls
exist if and only if $\beta>\frac{1}{4},$ which is the requirement that
$\phi=0$ is the unique global minimum of the potential.
In this case $\w_-<\w<\w_+,$ where the limiting values are
$\w_-=\sqrt{1-\frac{1}{4\beta}}$ and $\w_+=1.$ Both the charge $Q$ 
and energy $E$ are decreasing functions of $\w$ in this range, tending to
zero as the upper limit $\w_+$ is approached and growing without limit
as $\w\rightarrow \w_-.$ In Figure~\ref{fig-QE} the two upper curves
display the charge $Q$ (solid curve)
and energy $E$ (dashed curve) as a function of $\omega$ for $\beta=1/2,$
in which case $\w_-=1/\sqrt{2}.$

If $\beta\le\frac{1}{4}$ then there is an upper
bound on the value of $Q$ for which Q-ball solutions exist, as expected
in a theory where $\phi=0$ is not the unique global minimum of the potential.
In this case $w_-=0$ and $w_+=1,$ so the lower limit on the
frequency is removed, but Q-balls are unstable below a critical
value $\w\le\w_c,$ as we shall see shortly.
In Figure~\ref{fig-QE} the two lower curves
display the charge $Q$ (solid curve)
and energy $E$ (dashed curve) as a function of $\omega$ for 
the value $\beta=1/4.$

The properties of small Q-balls, which have arbitrarily small
values of $Q,$ were first investigated in \cite{Sp}, and 
are not very sensitive to the value of $\beta.$
Small Q-balls with $Q\ll 1$ correspond
to $\w$ close to the upper limit of the allowed range, 
$\omega\approx \w_+=1,$ that is $0<\w'\ll 1.$ In this limit
the above expressions for $Q$ and $E$ can be expanded in powers of 
$\w'$ to give
\be
Q=4\w'+\frac{1}{3}(16\beta -6)\w'^3+O(\w'^5),\quad
E=4\w'+\frac{1}{3}(16\beta -8)\w'^3+O(\w'^5).
\label{smallqe}
\ee
Clearly, the properties of small Q-balls only have a weak dependence upon
the value of $\beta,$ and the leading order behaviour is independent 
of $\beta.$ In fact this leading order behaviour is familiar from
the well-known limit of the Klein-Gordon equation to the nonlinear
Schr\"odinger equation for small amplitude breathers.

In this paper we shall be particularly concerned with small Q-balls,
and the above analysis shows that, at least to leading order,
the results are universal for all values of $\beta.$ 
We may therefore make a convenient choice and set $\beta=0,$ 
which we refer to as the truncated model. 
Taking the limit $\beta=0$ of the above results yields a simplified
expression for the profile function
\be
f=\w'\mbox{sech}(\w'x),
\label{trunprofile}
\ee
as well as the charge and energy
\be Q=4\w\w', \quad\quad
E=\frac{4\w'}{3}(1+2\w^2).
\label{trunqe}
\ee

It should be noted that the properties of small Q-balls will be 
well-approximated by the 
expressions in (\ref{trunprofile}) and (\ref{trunqe}) for the whole class
of theories where the potential has an expansion in $f^2$ of the form
\be
U(f)=m^2 f^2 -g^2 f^4+\ldots
\ee 
where $m$ and $g$ are positive constants which can be set to unity
by suitable rescalings of the field and spacetime coordinates.

Any model with small Q-balls will have the property that Q-balls tend
towards fundamental particles as the charge tends to zero, with the 
limit $E/Q\rightarrow m.$
The dispersion relation for a plane wave solution 
$\phi=\exp(i(\w t+kx))$ of the linearized theory is $\w^2=m^2+k^2,$ hence
$\w^2\ge m^2$ and this matches smoothly onto the end of the Q-ball 
frequency range $\w^2\le m^2$ with the exponential decay of the Q-ball
solution being interpreted as a complex wave number $k=i\sqrt{m^2-\w^2}.$ 

The condition for a Q-ball to be a classically stable solution is \cite{LP}
\be
\frac{dQ}{d\w}<0.
\ee 
For the truncated model ($\beta=0$) then $Q=4\w\w'$ and therefore a
Q-ball with $\w>\frac{1}{\sqrt{2}}=\w_c$ is stable. It is not surprising that
$\w_c=\frac{1}{\sqrt{2}}$ is the critical value at which instability emerges
since from (\ref{trunprofile}) this value corresponds to
 $f(0)=\frac{1}{\sqrt{2}},$
which is the turning point of the truncated potential
$U=f^2-f^4.$ The truncated model has a lower limit on the
charge, $Q\le 2,$ which contrasts with the theory with $\beta>\frac{1}{4},$
where stable Q-balls with arbitrarily large values of $Q$ exist.

\subsection{Q-balls in theories with a conformally flat target space}
An interesting generalization of the standard theory (\ref{usuallag})
is to consider a theory in which the complex field $\phi$ takes values 
in a conformally flat target space. Explicitly, the Lagrangian density
takes the form
\be
{\cal L}=\frac{1}{G(|\phi|)}\partial_\mu\phi\partial^\mu\bar\phi-W(|\phi|)\,,
\label{modlag}
\ee
where $G(|\phi|)$ is the inverse conformal factor of the target
space metric and $W(|\phi|)$ is the potential. Obviously the standard
Lagrangian (\ref{usuallag}) of the previous section is recovered by the choice 
 $G(|\phi|)=1$ and the identification $W(|\phi|)=U(|\phi|).$
Again we assume that the potential vanishes when the field is zero, that is,
$W(0)=0,$ and the metric is normalized by $G(0)=1.$

The conserved Noether charge in this theory is given by
\be
Q=i\int_{-\infty}^{\infty} 
\frac{1}{G}(\phi\partial_t\bar\phi-\bar\phi\partial_t\phi)\,dx.
\label{genq}
\ee
We shall mainly be concerned with the choice $G(f)=1-f^2,$ which has two 
main advantages. The first is that, as shown below, any potential (which
is finite for finite field values) that has a mass term yields Q-ball
solutions. This contrasts with the standard kinetic term which,
as discussed in the previous section, requires 
a rather careful choice of potential so that $U(f)/f^2$ has a minimum
at a positive value of $f.$  Of course, one may take the view that
the careful choice of $U$ has simply been replaced by a careful
choice of $G.$
The second advantage is that this choice
of $G$ includes an integrable theory, where exact solutions describing
multi-Q-ball dynamics can be constructed in closed form.

To begin with, Q-ball solutions will be discussed for arbitrary
functions $G$ and $W.$ This is most conveniently addressed by making
use of the momentum density, as follows.

The momentum density, $p,$ and the associated current $J^P$, which
follow from the Lagrangian density (\ref{modlag}), are given by
\be
p=-\frac{1}{G}(\partial_t\phi \partial_x\bar\phi
+\partial_t\bar\phi\partial_x\phi), \quad
J^P=-\frac{1}{G}(|\partial_t\phi|^2+|\partial_x\phi|^2)+W.
\label{modpj}
\ee
The conservation law $\partial_t p=\partial_x J^P$ implies that the
momentum $P$ is conserved, where
\be
P=\int_{-\infty}^{\infty}p\, dx.
\ee
A field of the Q-ball form (\ref{qball}) has $p=0$ and hence $J^P$
is constant. Therefore,
\be
J^P=-\frac{1}{G}\left(\w^2f^2+\left(\frac{df}{dx}\right)^2\right)+W=0,
\ee
where the constant value of $J^P$ is evaluated to be zero by
using the boundary condition $f\rightarrow 0$ as $x\rightarrow\infty.$

The profile function of the modified theory (\ref{modlag}) therefore
satisfies the first order equation \be
\left(\frac{df}{dx}\right)^2=-\w^2f^2+GW.
\ee
Note that this equation is identical to the profile function 
equation (\ref{bog})
after the identification $U=GW.$ 

With the choice $G(f)=1-f^2$ then
solutions exist for all positive $\w$ satisfying
\be
 \w^2<\w_+^2=\frac{1}{2}\frac{d^2W}{df^2}\bigg|_{f=0}
\label{conm1}
\ee
since the second constraint (\ref{con2}) with $U(f)=(1-f^2)W(f)$
is then automatically satisfied providing $W(1)$ is finite. 

The simplest theory with $G=1-f^2$ contains only a mass term,
which we normalize to be $W=f^2.$ This turns out to be an integrable
theory, and we discuss it in detail in the next section. Note that
in this case $U=GW=f^2-f^4,$ so the Q-ball solution is identical to that
of the truncated model with a standard kinetic term. 

Although we have restricted our discussion to theories in (1+1)-dimensions,
a similar analysis can be performed for the general theory (\ref{modlag})
in any number of space dimensions. The property that a non-standard
kinetic term, such as $G=1-|\phi|^2,$ removes the usual restrictions
on the potential to allow Q-balls is true in any dimension.
Non-standard kinetic terms are currently popular in cosmological
applications and arise in string theory, 
therefore there may be applications of Q-balls in this context 
to systems not previously considered.

\section{An integrable theory with Q-balls}\news
The integrable model referred to in the previous section 
is known as the complex sine-Gordon model \cite{LR,Ge} and
has a Lagrangian density given by
\be
{\cal L}=\frac{1}{1-|\phi|^2}\left(\partial_\mu\phi\partial^\mu\bar\phi
-|\phi|^2+|\phi|^4\right)\,,
\label{csglag}
\ee
where we have chosen to write all terms over a common denominator.
The numerator in (\ref{csglag}) is the Lagrangian density of the 
truncated model and the two theories share the same Q-ball profile function
(\ref{trunprofile}), for the reason described earlier.
Later, we shall be interested in studying small Q-balls in the complex 
sine-Gordon model, in which case $|\phi|\ll 1.$ However, it is not clear
which terms in the Lagrangian can be neglected in this limit, and the fact
that the profile function agrees with that of the truncated model (for all
$\w$, including small Q-balls) suggests that the dominant contribution
is obtained by approximating the denominator in (\ref{csglag}) by unity.
By comparing solutions we shall see later that this is indeed the case.
Note that simply keeping all terms to quadratic order in $\phi$ in
 (\ref{csglag}) does not provide a good description of small Q-balls
since such a theory does not allow Q-ball solutions.

Using the profile function (\ref{trunprofile}) the charge and energy
are calculated to be
\be
Q=4\cos^{-1}\omega,\quad\quad E=4\w'.
\ee
For small Q-balls the leading order behaviour is 
\be
Q=4\w'+O(\w'^3), \quad E=4\w',
\ee
which to first order in $\w'$ agrees with the result (\ref{smallqe})
for small Q-balls in the standard theory with potential (\ref{usualu}) 
and any value of $\beta.$

As $\w\rightarrow 0$ then naively $Q\rightarrow 2\pi,$ however, in this 
limit $f(0)\rightarrow 1$ and the integral which determines $Q$ is ill-defined
and has no unique limit. The integrable model therefore shares with the
truncated model the property that there is a maximum value of $Q,$ but
this is realized by two different kinds of behaviour in the two theories. 

The complex sine-Gordon model has been studied by a number of authors,
for example \cite{Ge},
and several different methods have been used to generate multi-soliton
solutions. However, it appears that the connection to Q-balls has not been
exploited previously, though it has been noted that solitons of this
theory are examples of Q-balls \cite{Mi}.
Interpretating the known multi-soliton solutions in the
framework of Q-balls and performing some analysis of these solutions will
prove instructive in understanding Q-ball dynamics and interactions in
non-integrable theories. In particular we shall make extensive use of the
two-soliton solution reproduced below.

Lorentz boosting the stationary Q-ball solution yields
\be
\phi_k={e^{i\theta_k} w_k'\mbox{exp}\bigg(iw_k\gamma_k(t-v_k(x-a_k))\bigg)}
\mbox{sech}\bigg(w_k'\gamma_k(x-a_k-v_kt)\bigg)\,,
\ee
where $\theta_k$ is an arbitrary constant phase, 
$\omega_k$ is the frequency, $\omega_k'$ is the complementary frequency,
$v_k$ is the velocity, with associated Lorentz factor 
$\gamma_k=1/\sqrt{1-v_k^2}$, and $a_k$ is the position of the Q-ball at
time $t=0.$

Defining the associated complex kink function
\be
\psi_k=-w_k'\mbox{tanh}\bigg(w_k'\gamma_k(x-a_k-v_kt)\bigg)-iw_k\,,
\label{complexkink}
\ee
and the constant 
\be
\delta_k=\sqrt{\frac{1-v_k}{1+v_k}},
\ee
then two Q-ball solutions $\phi_1,\phi_2,$ 
with associated kink functions $\psi_1,\psi_2$ can be combined to produce the 
two-soliton solution \cite{Ge}
\be
\phi=\frac{
(\delta_1\phi_1-\delta_2\phi_2)
(\delta_2\bar\psi_1-\delta_1\bar\psi_2)
-(\delta_1\phi_2-\delta_2\phi_1)
(\delta_2\psi_2-\delta_1\psi_1)}
{\delta_1^2-\delta_1\delta_2
(\phi_1\bar\phi_2+\bar\phi_1\phi_2+\psi_1\bar\psi_2+\bar\psi_1\psi_2)
+\delta_2^2
}.
\label{twosoliton}
\ee
Generically, this solution describes the scattering of two Q-balls,
with the traditional interpretation being that the two Q-balls 
pass through each other with their individual properties preserved 
and a time advance and phase shift being the only remnant of their 
interaction. 
In this and later sections we shall analyse this solution in some 
detail and see that it contains much more information than the above 
simple asymptotic picture might suggest. 

In contrast to most integrable systems there is no problem in setting 
both speeds $v_1$ and $v_2$ simultaneously to zero in this two-soliton 
solution. This yields the solution
\be
\phi=\frac{
ie^{i\w_1t}(\w_1-\w_2)\bigg(\w_1'
\mbox{sech}X_1
-e^{i(\w_2-\w_1)t+i\theta_2}\w_2'\mbox{sech}X_2\bigg)}
{1-\w_1\w_2-\w_1'\w_2'\bigg(\mbox{tanh}X_1
\mbox{tanh}X_2+\cos((\w_2-\w_1)t+\theta_2)\mbox{sech}X_1\mbox{sech}X_2
\bigg)},
\label{breather}
\ee
where we have defined 
$X_k=\w_k'(x-a_k),$ and without loss of generality we have set $\theta_1=0.$

The solution (\ref{breather}) describes a breather, that is, a non-stationary
solution which is periodic in time. The breather frequency is $|\w_2-\w_1|,$
and the solution degenerates to the trivial solution $\phi=0$ in the limit
$\w_1=\w_2.$ Breathers are usually associated with a soliton anti-soliton
bound state (for example, in the sine-Gordon model) but here the breather
is formed from two Q-balls, not a Q-ball anti-Qball configuration, though a 
breather solution of this form can also be obtained by considering 
negative values of $\omega_2.$ 

Given that the breather is constructed by
setting both velocities to zero in the two-soliton solution, then one
might expect that this solution describes two stationary 
Q-balls with fixed positions
and amplitudes. If the Q-balls are well-separated, that is,
the separation is much greater than either width, 
$|a_1-a_2|\gg \mbox{max}(\w_1'^{-1},\w_2'^{-1}),$ then this interpretation
is approximately true, although for any finite separation the position and 
amplitude of each Q-ball oscillates.  The amplitude of the oscillation is a
decreasing function of the separation between the Q-balls.

If both constituent Q-balls are placed at the same location (without
loss of generality this may be taken to be $a_1=a_2=0$) then it
describes an interesting oscillating periodic solution.
If $\w_1$ and $\w_2$ are comparable then the oscillation is between a state 
that resembles two resonably separated
Q-balls, each with charge $2\cos^{-1}\w_1+2\cos^{-1}\w_2,$ and an excited state
of a single Q-ball with charge $4\cos^{-1}\w_1+4\cos^{-1}\w_2.$ 
Note that because Q-balls are non-topological solitons then there is no
rigorous definition of the number of Q-balls in a given field configuration,
so it is perfectly reasonable for a field to have an appropriate
interpretation in which the number of Q-balls is different at various
times. 
In contrast, topological solitons have an associated conserved integer that
defines the number of solitons in any field configuration. Even though a 
topological multi-soliton solution may resemble a single structure,
the integer soliton number still allows its identification as a multi-soliton. 
If one of the frequencies is much closer to unity than the
other (say $\w_2\approx 1$) then the oscillation resembles a small
perturbation of a single Q-ball with frequency $\w_1.$ We shall see later
that this breather description of a perturbed Q-ball provides a good
explanation of observed Q-ball vibrational modes for non-integrable theories.
    
In the following section we investigate the force between
two well-separated Q-balls in a general theory. This will provide some
understanding of the fact that the breather solution requires the 
constituent Q-balls to have distinct frequencies.

\section{The force between Q-balls}\news
Numerical simulations of time-dependent nonlinear field theories
in one, two and three space dimensions, have revealed universal features
regarding the force between two Q-balls \cite{BS,AKPF}.
The numerical results show that two Q-balls with the same frequency can
attract or repel depending on whether their internal phases are aligned
or anti-aligned, and if the frequencies are distinct then a complicated
charge exchange process results. In this section we provide
an analytic calculation of the force between well-separated Q-balls
that explains this behaviour.

The force between two well-separated topological solitons is often
obtained by using a linear addition (or perhaps product) ansatz and calculating
the energy of this field as a function of soliton separation, in order to
derive the interaction potential. However, for Q-balls this method can not
be applied since the charge $Q$ of such an ansatz will vary with the 
separation, making a comparison of the energy a meaningless result. 
Explicitly, consider the addition ansatz
\be
\phi=\phi_1+\phi_2=e^{i\w_1t}f_1+e^{i\w_2t+i\theta}f_2
\label{add}
\ee
where $f_1=f_{\w_1}(x+a)$ and $f_2=f_{\w_2}(x-a)$ with
 $f_\w(x)$ the profile function for a Q-ball at the origin with 
frequency $\w.$ Here we choose $a$ to be positive and much greater than the
width of either Q-ball, so that the field (\ref{add}) is a good approximation
to two Q-balls with frequencies $\w_1$ and $\w_2$ at positions $-a$ and $a$
repsectively,
with $\theta$ the relative phase at $t=0.$  It is a simple task to verify
that the charge $Q$ of this field is not independent of $a,$ so computing the 
interaction energy as a function of $a$ is not a useful quantity. 

An alternative approach to calculating inter-soliton forces 
is to identify the force exerted on one soliton by the other with the induced
rate of change of momentum \cite{Ma5}. We now apply this approach to 
study the force between Q-balls in the general theory (\ref{modlag}),
which of course includes the standard theory (\ref{usuallag}) as
a special case. The force can also be calculated using the 
tail interaction methods of \cite{GO}.

Let $P[x_1,x_2]$ denote the momentum on the interval $[x_1,x_2]$, that is,
\be
P[x_1,x_2]=\int_{x_1}^{x_2}p\,dx=
-\int_{x_1}^{x_2}\frac{1}{G}(\partial_t\phi \partial_x\bar\phi
+\partial_t\bar\phi\partial_x\phi)\,dx.
\ee
The force on this interval, $F[x_1,x_2],$ is given by the time
derivative of the momentum, and the conservation law 
$\partial_t p=\partial_x J^P$ can be used to express this in terms of the
pressure difference at the end points
\be
F[x_1,x_2]=\frac{d P [x_1,x_2]}{dt}=\bigg[J^P\bigg]^{x_2}_{x_1}
=\bigg[-\frac{1}{G}(|\partial_t\phi|^2+|\partial_x\phi|^2)+W\bigg]^{x_2}_{x_1}.
\label{force1}\ee
We now apply the ansatz (\ref{add}) and note that the interval $[x_1,x_2]$ 
of interest will have end points which are far from both Q-balls, 
so that $\phi_1$ and $\phi_2$ are small where their evaluations are required,
 allowing the leading order asymptotic force to be calculated by keeping 
only terms up to quadratic order in $\phi.$ The functions
$G(|\phi|)$ and $W(|\phi|)$ in (\ref{force1}) can therefore be replaced
by $G=1$ and $W=|\phi|^2$ using the earlier normalizations. Using the 
fact that a single Q-ball has a vanishing current, the leading order
contribution to the force is given by
\bea
F[x_1,x_2]&=&\bigg[-\partial_t\phi_1\partial_t\bar\phi_2 
-\partial_t\bar\phi_1\partial_t\phi_2
-\partial_x\phi_1\partial_x\bar\phi_2 
-\partial_x\bar\phi_1\partial_x\phi_2
+ \phi_1\bar\phi_2 
+\bar\phi_1\phi_2\bigg]^{x_2}_{x_1}\nonumber \\
&=&
2\cos\big((\w_2-\w_1)t+\theta\big)\bigg[
(1-\w_1\w_2)f_1f_2-\frac{df_1}{dx}
\frac{df_2}{dx}
\bigg]^{x_2}_{x_1}.
\label{force2}
\eea
In order to identify the force on a soliton with the rate of change of
momentum on a large interval containing  this soliton, it is necessary
that there is no net flow of momentum in an interval which is far
from both solitons. For theories where a time-independent soliton exists
(which is the usual situation for topological solitons) this is automatic,
but Q-balls are non-topological solitons which are only stationary not static,
 and therefore this requirment is not guaranteed. 

To check the above requirement consider an interval around the origin $[-L,L]$ 
which is far from both solitons, that is, $a\gg L.$ The asymptotic tail 
of a Q-ball is given by 
\be
f_\w(x)\sim A_\w \exp(-\w'|x|),
\label{tail}
\ee
where $A_\w$ is a positive function of $\w.$ Therefore in the interval
 $[-L,L]$ we can use the tail approximations
\be
f_1\sim A_{\w_1}\exp(-\w_1'(x+a)),\quad
f_2\sim A_{\w_2}\exp(\w_2'(x-a)).
\label{tail2}
\ee
Substituting these expressions into (\ref{force2}) yields the result
\be
F[-L,L]=\cos\big((\w_2-\w_1)t+\theta\big)4A_{\w_1}A_{\w_2}
e^{-a(\w_1'+\w_2')}(1+\w_1'\w_2'-\w_1\w_2)\sinh((\w_2'-\w_1')L).
\label{force3}
\ee
This force is identically zero if and only if $\w_1=\w_2,$ therefore this is a
special case in which we can calculate the asymptotic force using the above
approach. For distinct frequencies the result (\ref{force3}) reveals
that there is a breather-like motion with a frequency $|\w_2-\w_1|.$
Indeed, we have seen that in the integrable model there is an exact
breather solution (\ref{breather}) with this frequency.

Restricting to the case of equal frequencies $\w_1=\w_2\equiv \w,$
gives $F[-L,L]=0$ and we can therefore 
calculate the force on the soliton at $x=a$
by computing $F[L,\infty],$ which will be independent of $L$ by the previous
calculation.

At $x=\infty$ there is clearly no contribution to the force because of the
exponential decay of both $f_1$ and $f_2.$ At $x=L$ we can again use
the asymptotic expressions (\ref{tail2}) but now with equal frequencies 
\be
f_1\sim A_{\w}\exp(-\w'(x+a)),\quad
f_2\sim A_{\w}\exp(\w'(x-a)).
\label{tail3}
\ee
Substituting these formulae into the expression (\ref{force2}) yields
\be
F\equiv F[L,\infty]=-2\cos\theta\bigg(
(1-\w_1\w_2)f_1f_2-\frac{df_1}{dx}
\frac{df_2}{dx}
\bigg)\bigg|_{x=L}
=-4\cos\theta \,A_\w^2\,\w'^2e^{-2a\w'},
\label{force4}
\ee 
which is indeed independent of $L,$ and has the expected exponential 
fall-off with separation $2a.$

For two in-phase Q-balls, $\theta=0$ and hence $F<0,$ so there is an attractive
force between the Q-balls, whereas if the Q-balls are exactly out-of-phase,
$\theta=\pi$ and hence $F>0$ and there is repulsion. This agrees with the
numerical simulations of the nonlinear field theory \cite{BS,AKPF}.
If $\theta$ is an integer multiple of $\pi$ then, by symmetry, two Q-balls
with equal initial frequencies will remain equal. 
However, for other values of $\theta$ there is no symmetry argument
to keep the initial frequencies equal, and indeed generically a breather-like
mode will evolve that induces a frequency difference, as discussed in
the next section. Therefore the force (\ref{force4}) can only be applied
to study the dynamics of Q-balls which are exactly in-phase or exactly
out-of-phase.  

For both the truncated model and the integrable model, the coefficient
$A_\w$ in the tail approximation (\ref{tail}) is given by $A_\w=2\w'.$
Hence in these two theories the asymptotic force is given by
\be
F=-16\cos\theta\,\w'^4e^{-2a\w'},
\label{forcet}
\ee
which is extremely weak for small Q-balls because of the $\w'^4$ factor;
recalling that the Q-ball mass approaches $4\w'.$

\begin{figure}
\begin{center}
\includegraphics[width=11cm]{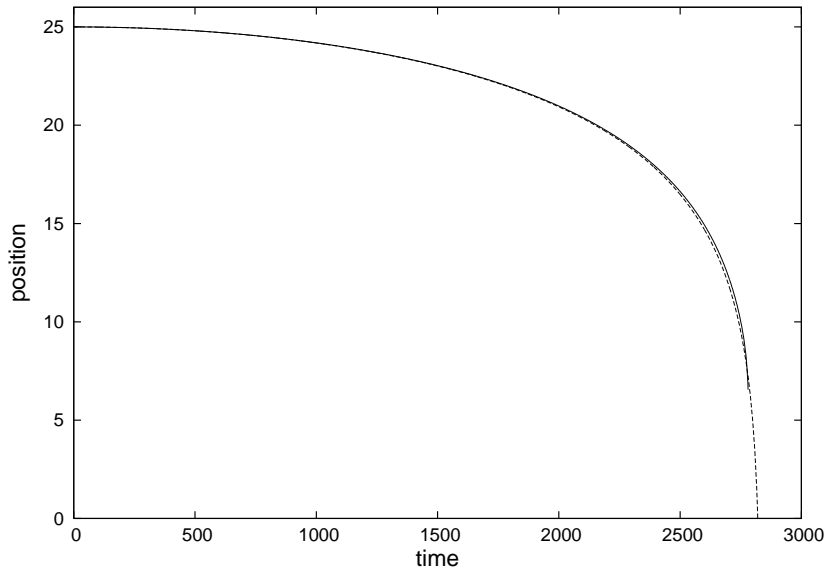}
\caption{The Q-ball position as a function of time, for an in-phase
 two-Q-ball system with equal frequencies $\w=0.98$  and
initial positions $\pm 25.$
The solid curve is the result of a numerical simulation and the dashed
curve is the approximation based on the asymptotic force. 
}
\label{fig-pos}
\end{center}
\end{figure}

The dynamics generated by the asymptotic force (\ref{forcet}) is 
obtained by solving the equation $M\ddot a=F,$ where the mass  $M$
is the energy of a stationay Q-ball with frequency $\w.$ 
Restricting to the truncated model then the mass is given by the second
expression in (\ref{trunqe}) and the equation of motion for two in-phase
Q-balls becomes
\be
\ddot a=-\frac{12\w'^3e^{-2a\w'}}{1+2\w^2}.
\ee
Taking initially stationary Q-balls, $\dot a(0)=0$, with $a(0)=a_0,$ 
the solution to this equation is
\be
a=a_0+\frac{1}{\w'}\log\bigg(\cos\bigg(
\frac{2\sqrt{3}\w'^2e^{-a_0\w'}t}{\sqrt{1+2\w^2}}
\bigg)\bigg).
\label{apos}
\ee
As an example, for the values $\w=0.98$ and $a_0=25,$ the position
(\ref{apos}) is plotted as the dashed curve in Figure~\ref{fig-pos}, for
times until the position vanishes. The solid curve on the same plot is
the result of a numerical simulation of the full nonlinear field theory
with the initial conditions formed using the addition ansatz
(\ref{add}) with the same parameter values. In the field theory simulation the 
position of each Q-ball is computed as the point at which $|\phi|$ is maximal.
Once the two Q-balls are sufficiently close together 
($a\approx 6.5$ in this example) the field $|\phi|$ no longer has two 
distinct maxima, but rather has a single maximum at the origin.
At this point the two Q-balls lose their individual identities and it
does not make sense to assign two Q-ball positions,
which is why the solid curve terminates in Figure~\ref{fig-pos}.
The evolution of the system beyond this point follows a damped breather
motion, as discussed in the following section.
There is excellent agreement between the two curves in Figure~\ref{fig-pos},  
even down to the smallest allowed separation. This demonstrates the accuracy
of the asymptotic force and reveals that it provides a good description
well beyond the expected region of validity.

Returning to the case of distinct frequencies, $\w_1\ne\w_2,$
then  although the force on a small interval far from both solitons 
(\ref{force3}) is non-zero, it does vanish if averaged over the time period  
$T={2\pi}/|\w_2-\w_1|.$ It would therefore make sense to compute
the time-averaged force on a Q-ball, but the expected leading order
contribution also averages to zero. We therefore see that two Q-balls
with distinct frequencies perform a breather-like motion where the
expected leading order contribution to the force averages to zero. In fact,
as we have seen, the integrable model has an exact breather solution,
hence in this theory the average force vanishes to all orders in the 
separation. For a generic theory one expects a higher order contribution
to be non-zero, which implies a very weak force between Q-balls.     

Note that the fact that the force does not vanish in the integrable
model if the two frequencies are equal explains why this limit 
degenerates in the exact breather solution.

Similar calculations to the ones performed in this section, to determine
the rate of change of momentum, reveal analogous results for the rate of 
change of charge. The breather-like motion is therefore also associated
with a quasi-periodic flow of charge and relates to the charge exchange
observed in numerical simulations \cite{BS}. This is discussed further in
the following section.  

\section{Breathers and multi-Q-ball dynamics}
\begin{figure}
\begin{center}
\includegraphics[width=10cm]{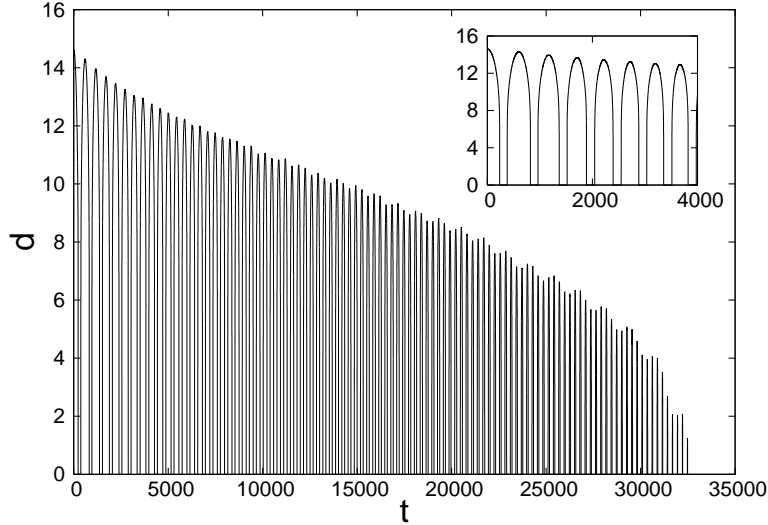}
\caption{The distance $d$ from the origin of the maximum of $|\phi|,$ 
for times $0\le t\le 35000.$ The simulation is performed in the truncated
model. The initial conditions (field and its time derivative at $t=0$) are
 taken from the exact breather solution
(\ref{breather}) of the integrable model, with
$\w_1=0.99,\ \w_2=0.98, \ \theta_2=\pi, \ a_1=a_2=0.$ The inset is a blow-up
of this plot for early times  $0\le t\le 4000.$ There are many oscillations
with a slowly decreasing amplitude. 
}
\label{fig-breather}
\end{center}
\end{figure}

\news
The exact breather solution (\ref{breather}) of the integrable model can
be used as an initial condition in the truncated model, to examine the 
differences between integrable and non-integrable evolution. The expectation
is that the breather motion is damped, as the non-integrable model 
radiates energy into other modes. A typical simulation is presented in
Figure~\ref{fig-breather} with the initial conditions obtained from the 
integrable solution with parameters
$\w_1=0.99,\ \w_2=0.98, \ \theta_2=\pi, \ a_1=a_2=0,$ corresponding
to a period of $T\approx 628.$ 
The graph displays the distance $d$ of either Q-ball from the origin,
where the Q-ball position is calculated as the location at which  $|\phi|$
is maximal. The inset graph shows the evolution for the first few
oscillations ($0\le t\le 4000$) revealing a slow decay of the amplitude, 
indicating that the maximal separation of the two Q-balls slightly decreases
with each cycle. Note that each oscillation contains a time interval where
the separation identically vanishes. During such an interval the maximal value
of  $|\phi|$ performs a single oscillation, but the maximal value  
remains located at the origin, that is, the configuration contains a single
peak rather than two peaks during this time interval. 

The main graph is a plot
for a much longer time period, $0\le t\le 35000,$ and confirms that many 
oscillations are performed before the system becomes a single Q-ball
configuration, in the sense that there is a single peak for $|\phi|$
for all time. In fact, a plot of the maximal value of $|\phi|$ as a function
of time shows that the configuration continues to oscillate, with a 
reasonably large variation in $|\phi|$ at the origin, for a length of time
which is far beyond that shown in Figure~\ref{fig-breather}. This part of the
evolution also has an approximate breather description, but with one 
of the frequencies much closer to unity than the other. Recall that in this 
parameter regime the exact breather solution of the integrable theory also
has the property that it remains a single peak throughout the entire period.
This will be analysed further when we discuss vibrational modes of a 
single Q-ball.

The dynamics described above may therefore be viewed from two different
perspectives. The first description is an oscillating two Q-ball system 
which dissipates energy until it becomes an oscillating single Q-ball and 
eventually a stationary single Q-ball. The second description is as a
slow drift through a family of breather solutions (labelled by 
the parameters $\w_1$ and $\w_2$) in which one frequency tends towards unity
while the other frequency decreases to approximately conserve charge.
Taking the limit $\w_2\rightarrow 1$ in the breather solution (\ref{breather})
yields the stationary single Q-ball solution with frequency $\w_1,$
so both descriptions are consistent. 

The single stationary Q-ball which results from the above evolution
has a charge which is a few percent less than the initial configuration.
The charge appears to be lost gradually, rather than in sudden drops, but this
is difficult to accurately measure numerically, particularly because it
is a small effect. Moreover, we are unable to present convincing evidence to
discriminate between the charge and energy being lost through radiation
or through the emission of tiny Q-balls. We suspect that the latter is  
correct, but because Q-balls tend towards elementary particles as the 
charge tends to zero, this makes it tricky to distinguish the two in a
noisy background.

In the numerical simulations described above the spatial domain has
a finite range. If Dirichlet boundary conditions are applied than
charge and energy are conserved throughout the simulation to a
very accurate precision. However, with such boundary conditions
energy and charge are observed to propagate to the boundary where
they are reflected back into the simulation domain. 
This is a standard phenomenon with Dirichlet boundary conditions
and does not accurately represent the true dynamics on an infinite
domain. To minimize the reflection produced by the boundary we apply
absorbing boundary conditions. This consists of a small region close
to the boundary where a damping term is introduced to dissipate most
of the energy and charge before it is reflected back into the
bulk of the simulation region. This provides a better approximation to
the true system on an infinite domain, and in particular allows charge
and energy to decrease, as it should for the theory defined on the
whole line when energy and charge are only computed in a finite interval. 

Similar results to the one described above are obtained for other 
parameter values. As the charge of the configuration increases the
damping becomes more significant and less oscillations are required until
a single peak forms. This is to be expected, since the integrable model is 
a better approximation to the non-integrable theory for smaller charge.
Note that the range of breathers that can be studied in the truncated
model is restricted, since stability requires that the field must 
satisfy the constraint $|\phi| < \frac{1}{\sqrt{2}}$ for all points in space
at all times. During a breather cycle the maximal value of $|\phi|$ can be
substantially larger than it is for either two distinct 
Q-balls or a single Q-ball, even when the configuration has the same charge
as the breather. 

The evolution of two equal frequency in-phase Q-balls, 
discussed in the previous section, also yields a damped breather motion, 
once the two Q-balls merge to
form a single peak for the first time. As the breather solution 
(\ref{breather}) degenerates for equal frequencies, the appropriate 
breather parameters, $\w_1$ and $\w_2$, to compare with 
(at least when the total charge
is small) are obtained by matching the total charge and energy of the
initial configuration to that of the breather.

\begin{figure}
\begin{center}
\includegraphics[width=15cm]{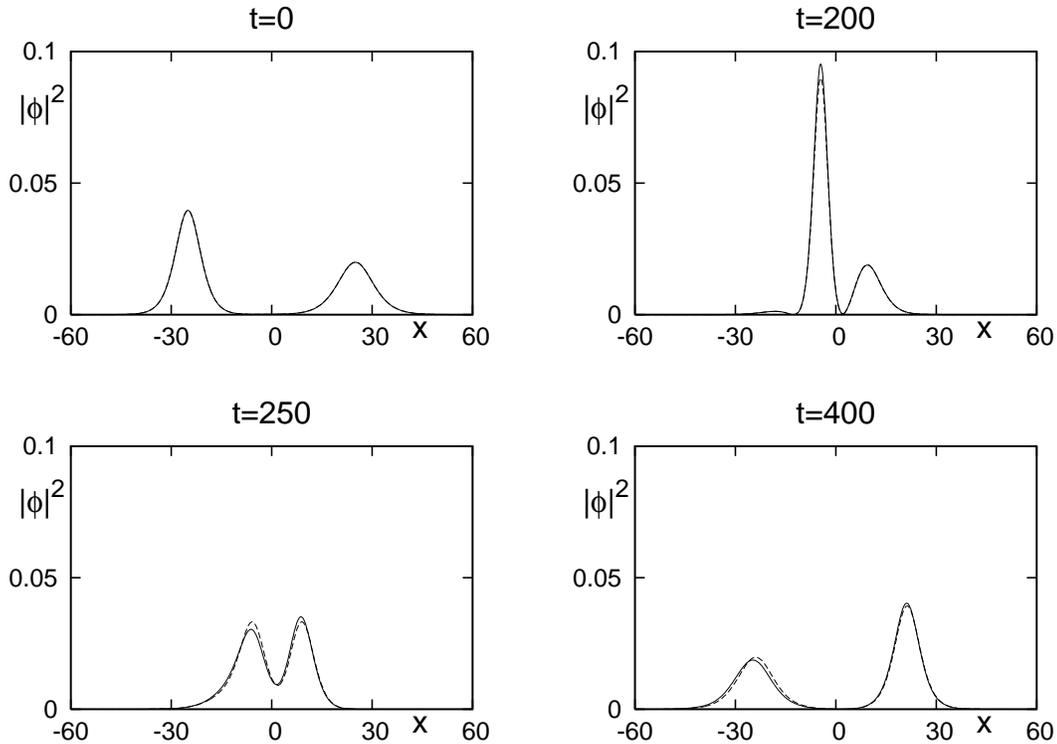}
\caption{
Plots of 
$|\phi|^2$ at increasing times in a two Q-ball scattering process
with frequencies $\w_1=0.98,$  $\w_2=0.99,$ initial positions 
$a_1=-a_2=-25,$ and speeds $v_1=-v_2=0.1.$
Solid curves are in the truncated model and dashed curves 
(often hidden by the solid curves) are in the integrable model.
}
\label{fig-scat}
\end{center}
\end{figure}

Recall that the two-soliton solution (\ref{twosoliton}) 
describes the elastic scattering of two Q-balls which 
emerge with their charges and speeds preserved. We therefore expect
the scattering of sufficiently small Q-balls to be close to this integrable 
behaviour for all systems in our generic class of theories. 

As an example, the solid curves in Figure~\ref{fig-scat} are plots of 
$|\phi|^2$ at increasing times in a two Q-ball scattering event in the 
truncated model. The initial condition is generated using the simple addition
ansatz with Q-balls of frequency $\w_1=0.98$ and $\w_2=0.99$ and respective 
positions $a_1=-a_2=-25.$  Both Q-balls are initially moving towards each
other with equal speeds, $v_1=-v_2=0.1,$ and are in-phase ($\theta=0$) at
$t=0.$ For comparison, the same scattering event in the integrable theory
is shown as the dashed curves in Figure~\ref{fig-scat} (though the two
evolutions are so close that it is difficult to distinguish the solid 
and dashed curves). Note that the exact solution (\ref{twosoliton}) 
could have been used to plot the dashed curves in Figure~\ref{fig-scat}, but
as a check on the numerical code and initial conditions, the integrable 
theory was also solved numerically -- reproducing the elastic scattering
result to a very high precision.

The simple asymptotic picture of trivial scattering in the integrable
theory hides the complicated interaction which takes place when the Q-balls
coalesce. During this part of the scattering process the configuration is 
highly distorted and the evolution resembles a breather-like motion.
In the integrable theory the two Q-balls emerge at exactly the point in 
the breather cycle at which the individual charges are identical to
the incoming charges. In the non-integrable theory the breather is slightly
distorted and the Q-balls emerge at a point in the cycle at which the
charges differ slightly from the initial values. With increasing charge
the breather cycle and the point of separation diverge between the integrable
and non-integrable models. This leads to a significant difference between
initial and final charges in non-integrable models, and results in the
charge exchange phenomenon seen in earlier numerical simulations \cite{BS}.

\section{Perturbing a Q-ball}\news
Perturbations of a single stable stationary Q-ball lead to extremely 
long-lived oscillations. In this section we discuss such oscillations
and provide formulae for the frequencies of the vibrational modes excited.

Given a Q-ball solution (\ref{qball}) of a standard theory (\ref{usuallag}),
we wish to consider perturbations which preserve the total charge (\ref{q}),
and therefore allow the perturbed solution to eventually return to the initial
unperturbed stationary Q-ball. A simple example is a modified squashing
perturbation, which uses an initial condition taken from the field 
configuration
\be \phi=\sqrt{\lambda}e^{i\w t}f(\lambda x).
\label{squash}\ee
If $\lambda=1$ then (\ref{squash}) is the stationary Q-ball solution,
but for other positive values of $\lambda$
 it describes an initial squashing or
stretching of the Q-ball. It is easily checked that the charge (\ref{q})
of this initial field configuration is
independent of $\lambda,$ and hence this is an example of a charge-preserving
perturbation.

\begin{figure}
\begin{center}
\includegraphics[width=10cm]{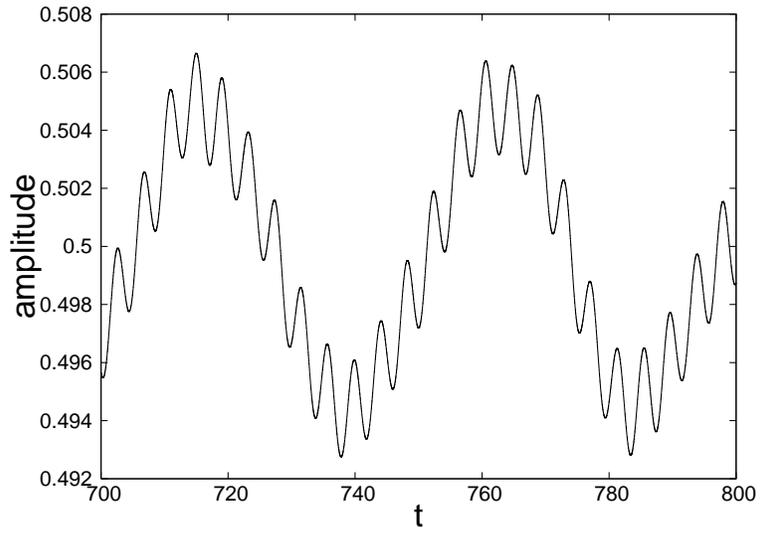}
\caption{The amplitude, that is $|\phi|$ at $x=0,$ as a function of
time $t\in[700,800]$, 
for a perturbed Q-ball with frequency $\omega=\frac{\sqrt{3}}{2}$
and squashing parameter $\lambda=1.05.$ 
}
\label{fig-pert}
\end{center}
\end{figure}

\begin{figure}
\begin{center}
\includegraphics[width=10cm]{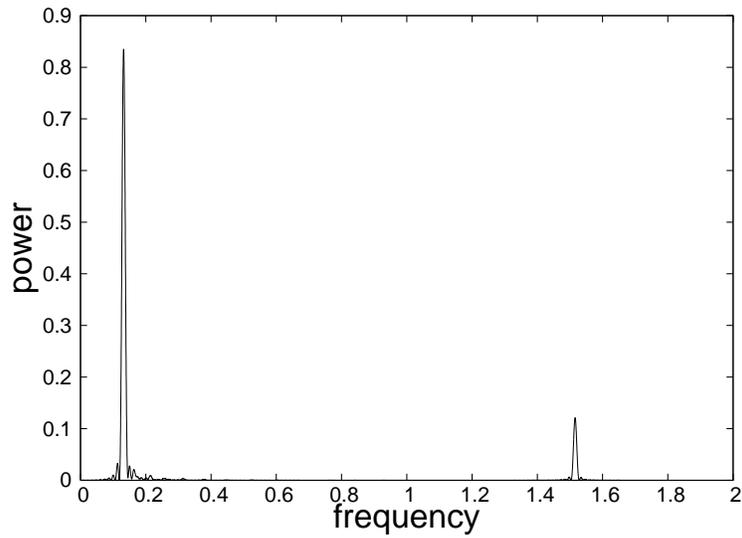}
\caption{The power spectrum of the oscillation 
displayed in Figure~\ref{fig-pert}.
}
\label{fig-power}
\end{center}
\end{figure}

Figure~\ref{fig-pert} displays part of the evolution for an initial
squashing of the above form with $\lambda=1.05$ and Q-ball frequency
$\omega=\frac{\sqrt{3}}{2}$ in the truncated model. The plot shows
the amplitude, that is $|\phi|$ at $x=0,$ for the time period 
$700\le t \le 800$. This demostrates the almost quasi-periodic nature of the
oscillation and the very slow decay of the amplitude. 

Figure~\ref{fig-power} is a plot of the power spectrum associated
with the oscillation presented in Figure~\ref{fig-pert}. 
It is clear that there are two dominant vibrational modes which are excited,
with frequencies $\Omega_B=0.132$ and $\Omega_F=1.516$ in this example.
Other charge-preserving perturbations yield similar results, as do 
scattering processess in which a Q-ball is excited through interactions
with other Q-balls. In each case two dominant vibrational modes are
excited and the aim in this section is to understand these modes and
obtain formulae for the frequncies $\Omega_B$ and $\Omega_F$ as a function
of the initial Q-ball frequency $\omega.$

Recall that the exact breather solution (\ref{breather}) of the integrable
theory has a frequency $|\omega_2-\omega_1|,$ where $\omega_1$ and $\omega_2$
are the frequencies of the two constituent Q-balls. Since the energy of
a Q-ball tends to zero as the frequency tends to unity, any arbitrary
perturbation of a single Q-ball contains enough energy to create an
additional Q-ball. Assuming the perturbation is small then any created
Q-ball must have a frequency close to unity. Taking $\omega_1=\omega,$
the frequency of the initial unperturbed Q-ball, and approximating
the frequency of the created Q-ball as $\omega_2\approx 1$ then we
may interpret the perturbed Q-ball as a breather with frequency
$\Omega_B=\omega_2-\omega_1\approx 1-\omega.$ 
In the numerical simulation presented above the values give $1-\omega=1-\frac{\sqrt{3}}{2}
=0.134\approx 0.132=\Omega_B.$ 

In summary, we see that the dominant mode excited in the perturbation of
a single Q-ball is a breather mode, with frequency 
$\Omega_B\approx 1-\omega.$ Other simulations of perturbed Q-balls
yield similar results and are consistent with this formula and its 
interpretation.
In particular this interpretation explains why the mode is so long-lived,
as it is close to a solution of the integrable theory which is strictly 
periodic. Note that if the perturbation is large then the energy imparted
to the created Q-ball can be significant, so the approximation that 
$\omega_2\approx 1$ may no longer be valid. In that case the frequencies
$\omega_1$ and $\omega_2$ should be calculated by matching the energy and 
charge of the perturbed Q-ball to the breather with parameters $\omega_1$
and $\omega_2.$ The result will be that $\omega_2$ is slightly less than
1 and $\omega_1$ is slightly greater than $\omega,$ in order to preserve
total charge. Thus the difference $\Omega_B$ will be slightly reduced
in comparison to the simple approximation      
$\Omega_B\approx 1-\omega.$ This has also been confirmed in numerical
simulations.

The interpretation of a perturbed Q-ball as a breather is similar
to a recent observation by K\"alberman \cite{Ka} on the oscillation
of a perturbed kink, called a wobble,
 in the sine-Gordon model. In that case a perturbed
kink can be described by an exact solution consisting of both a kink and a 
breather component. The difference for Q-balls is that the original
soliton is non-topological, so a closer analogy in the sine-Gordon theory 
would be the perturbation of a breather solution described by a two-breather
solution. The generic phenomenon is that any theory containing a 
non-topological soliton with arbitrarily small energy will produce such
a solution as a dominant low energy vibrational mode when any other
soliton (topological or non-topological) is perturbed. If the model is
integrable then an exactly periodic oscillation will result, but if the
theory is close to an integrable theory then the 
perturbation will yield a long-lived oscillation with a slowly decreasing
amplitude.
 
Now we turn to the second mode, with frequency $\Omega_F,$ observed
in the perturbation of a Q-ball. We shall show that this mode corresponds
to an oscillation in the internal frequency of the Q-ball by using 
the moduli space approximation (for a review see \cite{MS}) 
to determine the dynamics of the internal frequency by reducing the
field theory to a finite dimensional system.

Consider a field configuration of the form
\be
\phi=\alpha(t)e^{i\theta(t)}f(x),
\label{msfield}
\ee
where $\alpha$ and $\theta$ are time-dependent real moduli and 
$f(x)$ is the profile function of a Q-ball with frequency $\omega.$
Substituting the field (\ref{msfield}) into the Lagrangian of the truncated
model and performing the spatial integration yields the effective Lagrangian
\be
L=(\dot \alpha^2+\alpha^2\dot\theta^2)2\w'
-\alpha^22\w'+\frac{4}{3}\alpha^4\w'^3-\frac{2}{3}\alpha^2\w'^3.
\label{mslag}
\ee 
The equations of motion which follow from this Lagrangian are
\bea
& &\alpha\ddot\theta+2\dot\alpha\dot\theta=0=\frac{1}{\alpha}\frac{d}{dt}
(\alpha^2\dot\theta)\label{mseom1} \\
& &
\ddot\alpha-\alpha\dot\theta^2+\alpha(1+\frac{1}{3}\w'^2
-\frac{4}{3}\alpha^2\w'^2)=0.
\label{mseom2}
\eea
Equation (\ref{mseom1}) represents the conservation of Noether charge
associated with the field (\ref{msfield}). This reveals that at least two
degrees of freedom must be included in the moduli space approximation.
If the motion was reduced to only the internal phase $\theta$ then charge
conservation would force its velocity to be constant, and therefore
not allow an oscillation of the frequency $\dot\theta.$ Including the 
degree of freedom $\alpha$ allows a change in the amplitude to compensate
for a variation in the frequency.

The stationary Q-ball solution corresponds to $\dot\theta=\w$ and $\alpha=1,$
which is easily verified to be a solution of the equations (\ref{mseom1})
and (\ref{mseom2}). To study small oscillations around this solution
we set $\theta=\w t+\eta$ and $\alpha=1+\epsilon$ and linearize the
equations of motion in $\eta$ and $\epsilon$ to obtain
\be
\ddot\eta +2\w\dot\epsilon=0, \quad \mbox{and} \quad
\ddot\epsilon-2\w\dot\eta-\frac{8}{3}\w'^2\epsilon=0.
\label{mslin}
\ee
Integrating the first equation gives $\dot\eta=-2\w\epsilon+c,$
where $c$ is a constant determined by the initial perturbation. 
Substituting this into the second equation yields
\be
\ddot\epsilon+\frac{4}{3}(5\w^2-2)\epsilon-2\w c=0.
\ee
Hence the oscillation frequency is given by
\be
\Lambda=\frac{2}{\sqrt{3}}\sqrt{5\w^2-2},
\label{freqb}
\ee
which is automatically real since stable Q-balls only exist
in the truncated model for $\w^2>\frac{1}{2}.$

The earlier numerical simulation dealt with a Q-ball with frequency
$\w=\frac{\sqrt{3}}{2},$ which when substituted into the formula (\ref{freqb})
produces
\be
\Lambda=\sqrt{\frac{7}{3}}=1.528\approx 1.516=\Omega_F.
\ee
Perturbations of Q-balls for a range of values of $\w$ confirm that
the expression (\ref{freqb}) is a good approximation to the frequency 
$\Omega_F$ of the second observed vibrational mode. 
Furthermore, an initial perturbation of the form (\ref{msfield})
can be used as a charge-preserving perturbation in the full field theory
and a resulting power spectrum analysis shows that more of the energy
is imparted into the $\Omega_F$ mode than for other forms of perturbations, 
such as (\ref{squash}). 

The results of this section have demonstrated that a perturbation of
a stable Q-ball excites two main vibrational modes; the first a 
breather mode associated with the creation of a small Q-ball, 
and the second an oscillation of the internal frequency and amplitude of 
the Q-ball. Explicit formulae for the frequencies of these vibrations
have been derived for the truncated model and verified to be in good agreement
with numerical results. The same method can be used
to derive an analogue of the formula (\ref{freqb}) for the general
$\phi^6$ model with potential (\ref{usualu}) though the resulting expression
is more complicated.

\section{Duality}\news
It is known that the integrable complex sine-Gordon model (\ref{csglag})
has a dual description in which Q-balls of the original theory map
to kinks of the dual theory.  A reformulation of the theory
in terms of a gauged Wess-Zumino-Witten model reveals that this is
a T-duality relation \cite{PS}, and this has been 
extended to a range of integrable theories \cite{Mi}.
 In this section we show that a similar duality
transformation exists for any (1+1)-dimensional theory with
Q-balls and does not rely on the integrability of the theory.

We shall define a dual theory, with field $\psi,$ associated with the 
general theory (\ref{modlag}) with field $\phi$.
The first part of the construction requires a relation between
$|\psi|$ and $|\phi|$ given by solving the equation
\be
\frac{d|\psi|^2}{d|\phi|^2}=-\frac{|\psi|^2}{G(|\phi|)},
\label{deriv}
\ee
with the condition that $|\psi|=1$ when $|\phi|=0.$
Explicitly,
\be
|\psi|^2=\exp\int_0^{|\phi|^2}-\frac{df^2}{G(f)}.
\label{psiphi}
\ee
Given this relation we define the dual functions $\widetilde G(|\psi|)$
and $\widetilde W(|\psi|)$ via the formulae
\be
\widetilde G(|\psi|)=\frac{|\phi|^2|\psi|^2}{G(|\phi|)},
\quad\quad
\widetilde W(|\psi|)=W(|\phi|).
\label{dualgw}
\ee
The Lagrangian density of the dual theory is taken to be
\be
\widetilde{\cal L}=\frac{1}{\widetilde G(|\psi|)}
\partial_\mu\psi\partial^\mu\bar\psi-\widetilde W(|\psi|).
\label{duallag}
\ee
As an example, the complex sine-Gordon model has 
$G(|\phi|)=1-|\phi|^2$ and $W(|\phi|)=|\phi|^2.$
In this case
\be
|\psi|^2=\exp\int_0^{|\phi|^2}-\frac{df^2}{1-f^2}=1-|\phi|^2,
\ee
therefore
\be
\widetilde G(|\psi|)=\frac{|\phi|^2|\psi|^2}{G(|\phi|)}
=\frac{|\phi|^2|\psi|^2}{1-|\phi|^2}=1-|\psi|^2,
\ee
and
\be
\widetilde W(|\psi|)=W(|\phi|)=|\phi|^2=1-|\psi|^2.
\ee
The dual theory for the complex sine-Gordon model is then recovered
as 
\be
\widetilde{\cal L}=
\frac{1}{1-|\psi|^2}\left(\partial_\mu\psi\partial^\mu\bar\psi\right)
+|\psi|^2-1.
\label{dualcsg}
\ee
The complex kink field (\ref{complexkink}), used in constructing
multi-soliton solutions of the complex sine-Gordon model, is a solution
of the dual theory (\ref{dualcsg}), and we shall discuss this aspect more 
generally below. 

In the original theory the vacuum field upon which Q-balls are built is
$\phi=0$ and via (\ref{psiphi}) this maps to $|\psi|=1$ in the dual theory.
Therefore solutions of the dual theory are complex kinks which connect two 
different points on the unit circle as $x\rightarrow\pm\infty,$ with 
the phase difference in $\psi$ being a real-valued topological charge.
We shall denote this topological charge by ${\cal Q}$ and normalize
it as 
\be 
{\cal Q}=-2\{\arg(\psi(x=\infty))-\arg(\psi(x=-\infty))\}
=i\int\frac{\psi\partial_x\bar\psi-\bar\psi\partial_x\psi}{|\psi|^2}\,dx.
\label{topcharge}
\ee
Note that the value of $\psi$ at spatial infinity is fixed for dynamical
reasons since as $|x|\rightarrow\infty$ then $|\psi|\rightarrow 1$
 and therefore by (\ref{dualgw}) $G\rightarrow 0,$ freezing the dynamics
of the field at infinity.
 
In the following we describe how solutions of the dual theory are
obtained from those of the original theory. The relation (\ref{psiphi})
is used to obtain $|\psi|$ given $|\phi|$ but we still need to specify
how to obtain the phase of $\psi.$ 

The charge density $q$ of the conserved Noether charge $Q$ (\ref{genq})
in the original theory is
\be
q=
\frac{i}{G(|\phi|)}(\phi\partial_t\bar\phi-\bar\phi\partial_t\phi),
\ee
and the associated conservation law is $\partial_t q=\partial_x J^Q$
where
\be
J^Q=\frac{i}{G(|\phi|)}(\phi\partial_x\bar\phi-\bar\phi\partial_x\phi).
\ee
This allows the definition of a field $\Theta$ through the relations
\be
\partial_x\Theta=\frac{1}{2}q, \quad \mbox{and}\quad  \partial_t\Theta=\frac{1}{2}J^Q.
\label{Theta}
\ee 
We now prove that 
\be
\psi=e^{-i\Theta}|\psi|,
\label{defnpsi}
\ee
is a solution of the dual theory.

Combining (\ref{defnpsi}) with (\ref{Theta}) gives
\be
\frac{\bar\psi\partial_x \psi-\psi\partial_x\bar\psi}{|\psi|^2}
=\frac{\phi\partial_t\bar\phi-\bar\phi\partial_t\phi}{G(|\phi|)},
\quad\mbox{and}\quad 
\frac{\bar\psi\partial_t \psi-\psi\partial_t\bar\psi}{|\psi|^2}
=\frac{\phi\partial_x\bar\phi-\bar\phi\partial_x\phi}{G(|\phi|)}.
\label{rel1}
\ee
Calculating $\partial_\mu|\psi|^2$ and using (\ref{deriv}) to express
this in terms of $\phi$ yields
\be
 \frac{\bar\psi\partial_\mu \psi+\psi\partial_\mu\bar\psi}{|\psi|^2}
=-\frac{\phi\partial_\mu\bar\phi+\bar\phi\partial_\mu\phi}{G(|\phi|)}.
\label{rel2}
\ee
Equations (\ref{rel1}) and (\ref{rel2}) are most conveniently expressed
using light cone coordinates $u=\frac{1}{2}(t+x)$ and $v=\frac{1}{2}(t-x)$
to give
\be
\frac{\bar\psi\partial_u\psi}{|\psi|^2}
=-\frac{\bar\phi\partial_u\phi}{G(|\phi|)},
\quad\mbox{and}\quad 
\frac{\bar\psi\partial_v\psi}{|\psi|^2}
=-\frac{\phi\partial_v\bar\phi}{G(|\phi|)}.
\label{lcrel}
\ee
In light cone coordinates the field equation which follows from
the dual Lagrangian density (\ref{duallag}) is 
\be
\partial_{uv}\psi-\partial_u\psi\partial_v\psi
\frac{\bar\psi}{\widetilde G(|\psi|)}
\frac{d\widetilde G(|\psi|)}{d|\psi|^2}
+\psi\widetilde G(|\psi|)\frac{d\widetilde W(|\psi|)}{d|\psi|^2}
=0.
\label{lcdual}
\ee
Using the expressions in (\ref{lcrel}) together with (\ref{deriv})
and the definitions (\ref{dualgw}) for $\widetilde G$ and $\widetilde W$
 it can be shown that
\bea
& &\partial_{uv}\psi-\partial_u\psi\partial_v\psi
\frac{\bar\psi}{\widetilde G(|\psi|)}
\frac{d\widetilde G(|\psi|)}{d|\psi|^2}
+\psi\widetilde G(|\psi|)\frac{d\widetilde W(|\psi|)}{d|\psi|^2}\\
&=&
-\frac{\psi\bar\phi}{G(|\phi|)}
\left\{
\partial_{uv}\phi-\partial_u\phi\partial_v\phi
\frac{\bar\phi}{G(|\phi|)}
\frac{d G(|\phi|)}{d|\phi|^2}
+\phi G(|\phi|)\frac{d W(|\phi|)}{d|\phi|^2}\right\}.
\eea
The vanishing of the final expression in the above is equivalent
to the field equation of the original theory in light cone coordinates,
hence we have proved that the definition (\ref{defnpsi}) with
(\ref{psiphi}) produces solutions of the dual theory from solutions
of the original theory.

As $-\Theta$ is the phase of $\psi$ the first relation in (\ref{Theta})
shows that the topological charge (\ref{topcharge}) in the dual 
theory is equal to the Noether charge in the original theory since
\be
{\cal Q}=2\big[\Theta\big]^{x=\infty}_{x=-\infty}=2\int^\infty_{-\infty}
\partial_x\Theta\, dx= \int^\infty_{-\infty}q\, dx=Q.
\ee
A stationary Q-ball in the original theory has $J^Q=0,$ therefore by
the second relation in (\ref{Theta}) $\Theta$ is independent of time.
Therefore we have shown that in any theory a stationary Q-ball solution
 has a  description in a dual theory as a static kink, with an interchange
of Noether and topological charges.
It can also been shown that the energy and momentum is preserved under duality
and that the Lagrangian densities of the original and dual theories 
are equal up to a total derivative. 

The dual theory (\ref{dualcsg}) for the complex sine-Gordon model
has kink solutions with a continuous range of energies,
because the vacuum points at spatial infinity are any two different 
points on the unit circle. It is interesting
that embedding these kinks as domain walls
in a higher dimensional theory therefore yields domain walls with
a continuous range of tensions and allows the construction of
BPS junctions with arbitrary angles \cite{NNS}.  
 
It remains to be seen whether the
dual description of Q-balls can be exploited to further investigate 
the dynamics and interactions of Q-balls, but it may prove useful as
static kinks are generally easier to deal with than Q-balls, because
the complication due to time dependent phases does not arise.

It is interesting to note that there is a self-dual theory defined by
\be
{\cal L}=
\frac{1}{1-|\phi|^2}\left(\partial_\mu\phi\partial^\mu\bar\phi\right)
-|\phi|^2+|\phi|^4,
\label{csgII}
\ee
and is another known example of an integrable theory,
generally referred to as the complex sine-Gordon II model \cite{Ge2}.
This model has degenerate vacua at $|\phi|=0$ and $|\phi|=1.$  
From our earlier analysis the integrable model (\ref{csgII}) 
shares the same Q-ball solutions
as the standard theory (\ref{usuallag}) with the potential
$U(|\phi|)=|\phi|^2(1-|\phi|^2)^2.$ The integrable model (\ref{csgII})
could therefore be used to study this theory, in the same way that the
integrable model (\ref{csglag}) was used to study the truncated theory. 
However, as we have already demonstrated, the properties of small Q-balls
in both integrable theories will converge to the same limit as the charge
decreases (once the appropriate rescalings have been applied to match
normalizations).

\section{Conclusion}\news
In this paper we have performed some of the first analytic studies
of Q-ball dynamics and interactions. We have shown how an integrable
theory can be used to study small Q-balls in non-integrable theories
and explain some of the phenomena observed in previous numerical
investigations. Although most of our analysis has been applied to small Q-balls,
and has been restricted to (1+1)-dimensions, there appears to be much in
common with numerical simulations in higher dimensions and for larger Q-balls
\cite{BS,AKPF}. This suggests that the results should be of some relevance
to more general issues regarding Q-balls. One obvious difference is that
in (3+1)-dimensions a theory with a standard kinetic term and a
 potential which is polynomial in $|\phi|^2$ does not allow small Q-balls,
although a potential which is polynomial in $|\phi|$ does have small
Q-balls \cite{K1}. This difference can be understood by examining 
the conditions required for the existence of small Q-balls for a general 
potential in an arbitrary number of space dimensions \cite{TCS}.

Finally, the integrable theory also provides an effective approximation 
to small Q-ball anti-Qball dynamics. As the integrable theory has an exact
periodic solution of this type then, as expected, in non-integrable
theories the behaviour is a perturbation of this periodic case. The
Q-ball and anti-Qball oscillate together for a number of cycles and
the annihilation process is slow, with only a small amount
of charge lost in each cycle. The fact that Q-ball anti-Qball annihilation
is a slow and inefficient process could have repercussions in the cosmological
context of Q-ball formation in phase transitions.

\section*{Acknowledgements}
DF thanks the STFC for a research studentship.
PMS thanks Steve Abel and Nick Manton for useful discussions,
and is grateful to the hospitality of Dionisio Bazeia and the
Federal University of Paraiba, Brazil, where this manuscript was completed.
We thank Luis Miramontes for drawing our attention to the paper \cite{Mi}.
The numerical computations were performed on the Durham HPC cluster HAMILTON.


\begin{thebibliography}{100}

\bibitem{AD} I. Affleck and M. Dine,
\textit{Phys. Lett.} \textbf{B249}, 361 (1985).

\bibitem{AKPF}
M. Axenides, S. Komineas, L. Perivolaropoulos, M. Floratos,
\textit{Phys. Rev.} \textbf{D61}, 085006 (2000).

\bibitem{BS} R.~A. Battye and P.~M. Sutcliffe,
\textit{Nucl. Phys.} \textbf{B590}, 329 (2000).

\bibitem{Co} S. Coleman,
\textit{Nucl. Phys.} \textbf{B262}, 263 (1985).

\bibitem{LR} F. Lund and T. Regge,
\textit{Phys. Rev.} \textbf{D14}, 1524 (1976).

\bibitem{Ge} B.~S. Getmanov,
\textit{Zh. Eksp. Teor. Fiz.} \textbf{25}, 132 (1977).

\bibitem{Ge2} B.~S. Getmanov,
\textit{Teor. Mat. Fiz.} \textbf{48}, 13 (1981).

\bibitem{GO} K.~A. Gorshkov and L.~A. Ostrovsky,
\textit{Physica} \textbf{D3}, 428 (1981).

\bibitem{Ka} G. K\"alberman,
\textit{J. Phys.} \textbf{A37}, 11607 (2004).

\bibitem{K1} A. Kusenko,
\textit{Phys. Lett.} \textbf{B404}, 285 (1997).

\bibitem{K2} A. Kusenko,
\textit{Phys. Lett.} \textbf{B405}, 108 (1997).

\bibitem{LP} T.~D. Lee and Y. Pang,
\textit{Phys. Rept.} \textbf{221}, 251 (1992).

\bibitem{Ma5} N.~S. Manton,
\textit{Nucl. Phys.} \textbf{B150}, 397 (1979).

\bibitem{MS} N.~S. Manton and P.~M. Sutcliffe, {\sl Topological Solitons},
Cambridge University Press (2004).

\bibitem{Mi} J.~L. Miramontes,
\textit{Nucl. Phys.} \textbf{B702}, 419 (2004).

\bibitem{NNS} M. Naganuma, M. Nitta, N. Sakai,
\textit{Phys. Rev.} \textbf{D65}, 045016 (2002).

\bibitem{PS} Q-Han Park and H.~J. Shin,
\textit{Phys. Lett.} \textbf{B359}, 125 (1995).

\bibitem{Sp} D. Spector,
\textit{Phys. Lett.} \textbf{B194}, 103 (1987).

\bibitem{TCS} M.~I. Tsumagari, E.~J.Copeland and P.~M. Saffin,
\textit{Phys. Rev.} \textbf{D78}, 065021 (2008). 

\end{thebibliography}
\end{document}